\documentclass[manuscript]{acmart}
\setcopyright{none}

\usepackage{graphicx}
\usepackage{booktabs}
\usepackage{tabularx}
\usepackage{array}
\usepackage{enumitem}
\usepackage{float}
\usepackage{multicol}
\usepackage{hyperref}
\usepackage{wrapfig}

\hypersetup{
  colorlinks=true,
  linkcolor=blue,
  citecolor=blue,
  urlcolor=blue
}

\newcolumntype{Y}{>{\raggedright\arraybackslash}X}
\newif\ifcacm
\cacmfalse
\newif\ifarxiv
\arxivtrue

\newcommand*{\WrapLR}[3]
{\bgroup
  \sbox0{\raisebox{-\height}{#1}}
  \sbox1{\raisebox{-\height}{#2}}
  \sbox2{\raisebox{\depth}{\makebox[\textwidth]{\usebox0\hfill\usebox1}}}
  \par\noindent\usebox2\vspace{\dimexpr \ht\strutbox-\ht2-\baselineskip-\parskip}\par
  \dimen0=\dimexpr \wd0+\columnsep\relax
  \dimen1=\dimexpr \textwidth-\wd0-\wd1-2\columnsep\relax
  \edef\shape{\the\dimen0 \the\dimen1}
  \dimen2=\ht\strutbox
  \count1=2
  \loop\ifdim\dimen2<\ht2
    \ifdim\dimen2>\dp0\relax
      \dimen0=0pt
      \dimen1=\dimexpr \textwidth-\wd1-\columnsep\relax
    \fi
    \ifdim\dimen2>\dp1\relax
      \dimen1=\dimexpr \textwidth-\wd0-\columnsep\relax
    \fi
    \advance\dimen2 by \baselineskip
    \advance\count1 by 1
    \edef\shape{\shape\space\the\dimen0 \the\dimen1}%
  \repeat
  \edef\shape{\the\count1 \space\shape\space 0pt \the\textwidth}%
  \parshape=\shape
  #3\par
\egroup}

\title{I hope we don't do to trust what advertising has done to love}

\author{Jade Alglave}
\affiliation{%
  \institution{Arm and University College London}
  \country{United Kingdom}}
\email{jade.alglave@arm.com} \email{j.alglave@ucl.ac.uk}

\renewcommand{\shortauthors}{Jade Alglave}

\begin{abstract}
Advertising uses love to sell stuff, like nylons. It also uses
the word ``love'' in trivialising ways---do you ``love'' your oven?
When I hear about trust in the context of AI, especially agentic, I hope we
don't do to trust what advertising has done to love.

But what is trust? Can we discuss it in actionable and measurable ways in the
context of AI? Thus I suggest a number of ``trust pillars'', hoping to start a
communal conversation, across computing and beyond, to civil society. I also
suggest that agentic systems may be a blessing in disguise, as we may be able
to turn their explicit interfaces into ``trust vectors''.
\end{abstract}


\begin{document}
\maketitle

\ifarxiv
\WrapLR{\includegraphics[height=.8in]{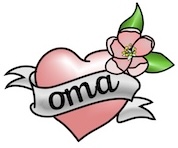}}%
{\includegraphics[height=.8in]{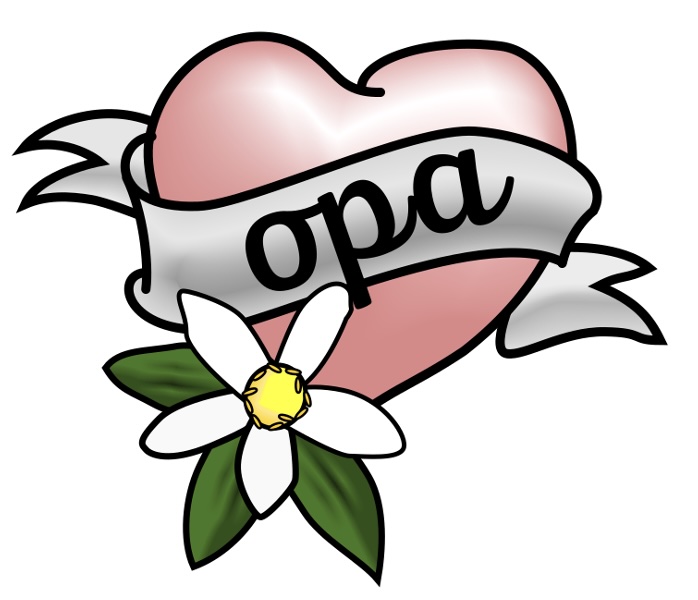}}%
{\noindent 
My grandma is not tech-savvy: I worry about what technology looks like and does
to her; especially if AI, whether agentic or not, becomes a prevalent way to
interact with various services she depends on.
How is she to trust that it's
done what it was supposed to, and nothing else? 
Maybe you worry about your folks in that way too. This column is for all of them.
}
\else
\noindent
My grandma is not tech-savvy: I worry about what technology looks like and does
to her; especially if AI, whether agentic or not, becomes a prevalent way to
interact with services she depends on.
How is she to trust that it's
done what it was supposed to, and nothing else?
Maybe you worry about your folks in that way too. This column is for all of them.
\fi


\section{AI is funny} AI applications may give absurd answers. These
applications progress fast: the number of absurd answers may decrease. But we
still don't really know why they behave how they do~\cite{woo26}\ifarxiv\cite{mol26}\fi.

Yet the AI world marches forward, towards agentic systems: AI applications that
don't just chat back at you, but can do stuff on your behalf. I hear a lot
about this making the question of trust worse, so it got me thinking. What are
agentic systems? Might they be a blessing in disguise? What is ``the question
of trust''? What agentic systems are we looking to trust? How can we architect
such systems? Would this be enough for my grandma?

\section{What are agentic systems?}

\begin{figure}[!h]
\centering
\begin{minipage}[t]{0.48\linewidth}
\centering
\includegraphics[width=\linewidth]{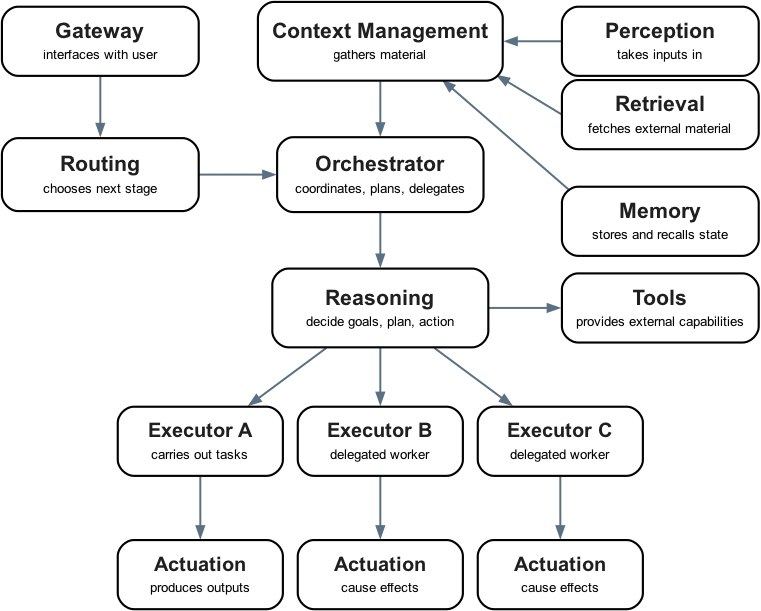}
\end{minipage}\hfill
\begin{minipage}[t]{0.48\linewidth}
\centering
\includegraphics[width=\linewidth]{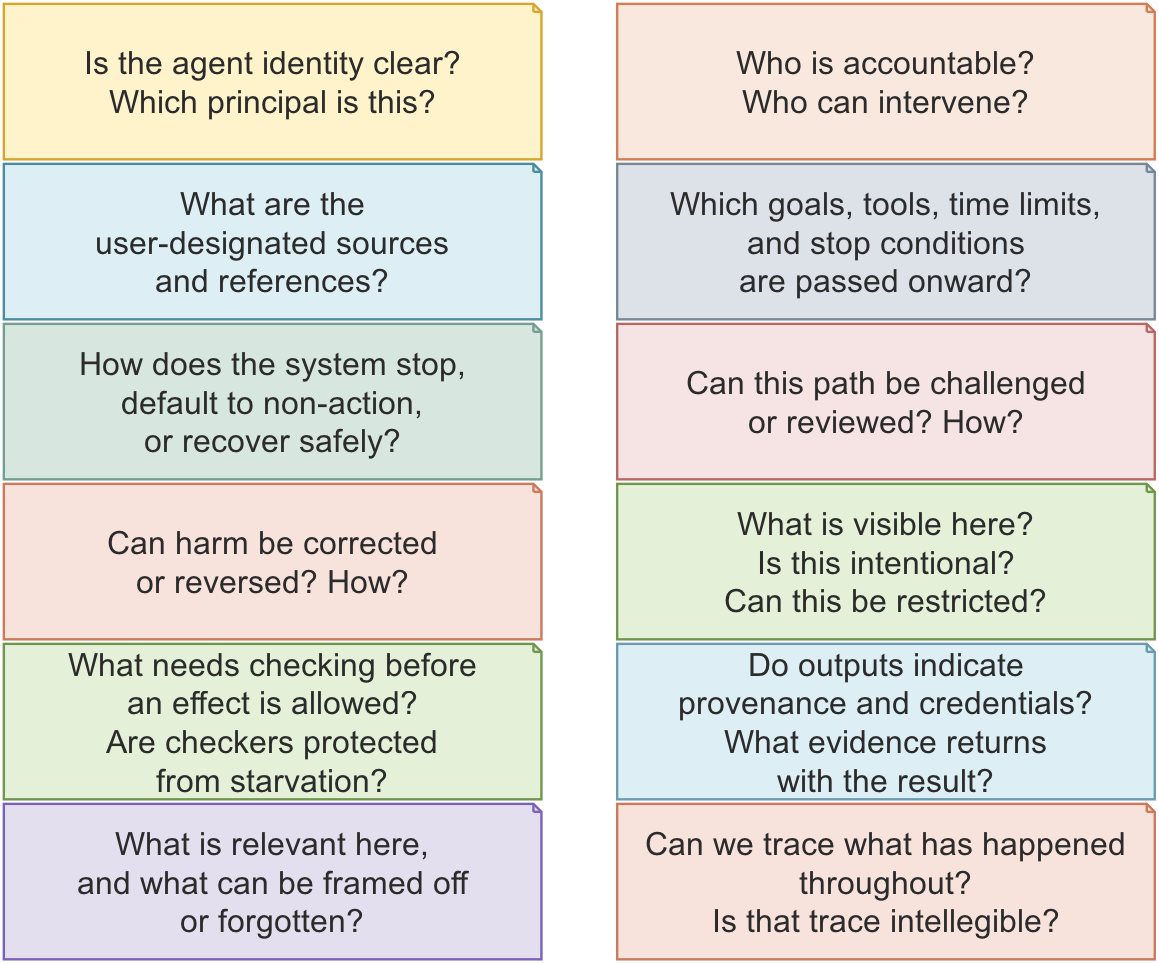}
\end{minipage}
\caption{Basic agentic architecture and Examples of trust-related questions \label{fig:basic-agentic-shape}}
\ifcacm
\vspace*{-10mm}
\fi
\end{figure}

In first approximation, a system is agentic if an LLM can influence what it
does next.  More concretely, I mean a system as in
Figure~\ref{fig:basic-agentic-shape}: there, the ``Reasoning'' box at least,
and perhaps others, may often be an LLM. But you can see that we are now
examining not just a single black-box LLM but a whole system around it. 

Back to my grandma, say she needs to renew her passport via an AI interface and
part way through the doorbell rings: Gateway interfaces through her phone,
Routing chooses the next stage, Context Management gathers relevant material,
e.g. a snapshot of her old passport, Orchestrator coordinates, plans,
delegates, Reasoning proposes next steps, Perception notices the doorbell,
Retrieval fetches external material e.g. guidance on what picture is
acceptable, Memory stores and recalls state, e.g.  where she was before
answering the door, Tools provide external capabilities e.g. her phone camera,
Executors carry out tasks e.g.  filling in the form, and Actuation causes
effects, finally filing her application.

Let's distinguish model execution and surrounding system execution. The model
produces e.g. a tool selection, a ranking, a plan; the surrounding system uses
the model, accesses memory, manages the context, calls the tool.  Components in
the agentic architecture may not need to run sequentially, but a good
understanding of what can run concurrently is not apparent.  Components don't
need to all be located in the same place, or be static: Executors may be
elsewhere than a local Orchestrator, some amount of Retrieval, Context
Management, or Reasoning may be remote, etc.
 
\section{Might agentic systems be a blessing in disguise?}

Thus agentic systems may alter visibility, control flow, authority unduly. But
the shift to an agentic architecture is an opportunity; we can ask: what
is visible? Is that intentional? What needs checking ahead of effects? What is
relevant, can or must be framed off or forgotten? Is someone accountable for
the effects? Can they intervene? Can my grandma request redress?... I give a
few more questions in Figure~\ref{fig:basic-agentic-shape}.

If architected well, agentic systems may be more trustworthy than a black-box
LLM, as they expose interfaces. We could e.g.: intentionally restrict
visibility before retrieval; soundly, e.g. programmatically or symbolically,
reduce context; carry sources and credentials throughout execution; delineate
and enforce delegation limits; run checks before taking action.  Trust might
then be less about the LLM alone and more about the architecture of the
end-to-end system. 

As an aside, and acknowledging Richard Sutton’s ``Bitter
Lesson''\ifarxiv~\cite{sut19}\fi, I wonder: could we do more ``on the cheap''?
Can we reproduce, or get close enough to\ifarxiv~\cite{odl19}\fi, results obtained with
LLMs, with maybe small, maybe domain-specific, maybe open, models wrapped in
carefully crafted agentic frameworks? See e.g.~\ifarxiv\cite{bhd25}\fi\cite{aisle-vs-mythos}.

Might cheap, small, domain-specific or open models be easier both to check
and keep in check?  Frontier LLMs may require Fort Knox, whilst other models
might remain politely within a small hotel safe.  Safety-critical systems may
prefer simpler and better understood architectures over maximal capability:
historically, avionics has been cautious about adopting complex systems such as
multiprocessors and set up safety nets and checks around them. 

Overall, we need to understand and document what LLMs can and cannot do with
purposeful, ambitious, rigorous, reproducible and open studies: how do they
fare on termination\ifarxiv~\cite{ohearn-halting}\fi?  Can they find concurrent
invariants in Linux?... 

And whilst for commercial reasons, implementation transparency may be unavailable,
can we provide extensional definitions? Arm has demonstrated the technical and
commercial feasibility of devising, maintaining and publicly distributing a
behavioural envelope for systems as complex as multiprocessors, without
revealing, or indeed prescribing, any implementation insight\ifarxiv~\cite{arm-arm}\fi.
Might this be achievable for LLMs? 

What architecture is needed to sustain these ``close enough'' agentic
frameworks reliably? What compute component resides where, presumably in a
concurrent, distributed, layered, dynamic fashion across e.g.\ personal devices
and remote locations? Where do verification components reside? Can we make sure
they're not neglected in favour of performance?

\section{What is ``the question of trust''?}

I found this definition: ``Trust is a willingness to open oneself to risk by
engaging in a relationship with another party''~\cite{hon-grunig}. Trust is not
in our control, but granted by the user: we must have user scenarios in any
trust-related proposal. 

My grandma needing to renew her passport is one, but we need more. A personal
agent buys a plane ticket: will it drain your savings? A decision procedure for
recruitment, automatically ranking candidates: will it follow policies, avoid
bias\ifarxiv~\cite{ucw20}\fi? An AI tutor for children, supporting homework: will it
follow the curriculum? Facial recognition to curb shoplifting: will it
systematically flag the same demographics?\ldots  

We can at least try to offer users justification for relying on a system:
a ``trustworthy entity is one for which sufficient evidence exists to support
its claimed trustworthiness''~\cite{nist}. This justification depends on
context---country, culture\ldots\ifarxiv~\cite{melbourne}\fi---e.g AI-aided
farming can't be one-size-fits-all when landscapes differ between Marfa and
Osaka.

The trust ``pillars'' in Figure~\ref{fig:trust-pillars} attempt to make
potential elements of justification explicit.

\ifarxiv
\begin{figure}[!h]
\includegraphics[width=\linewidth]{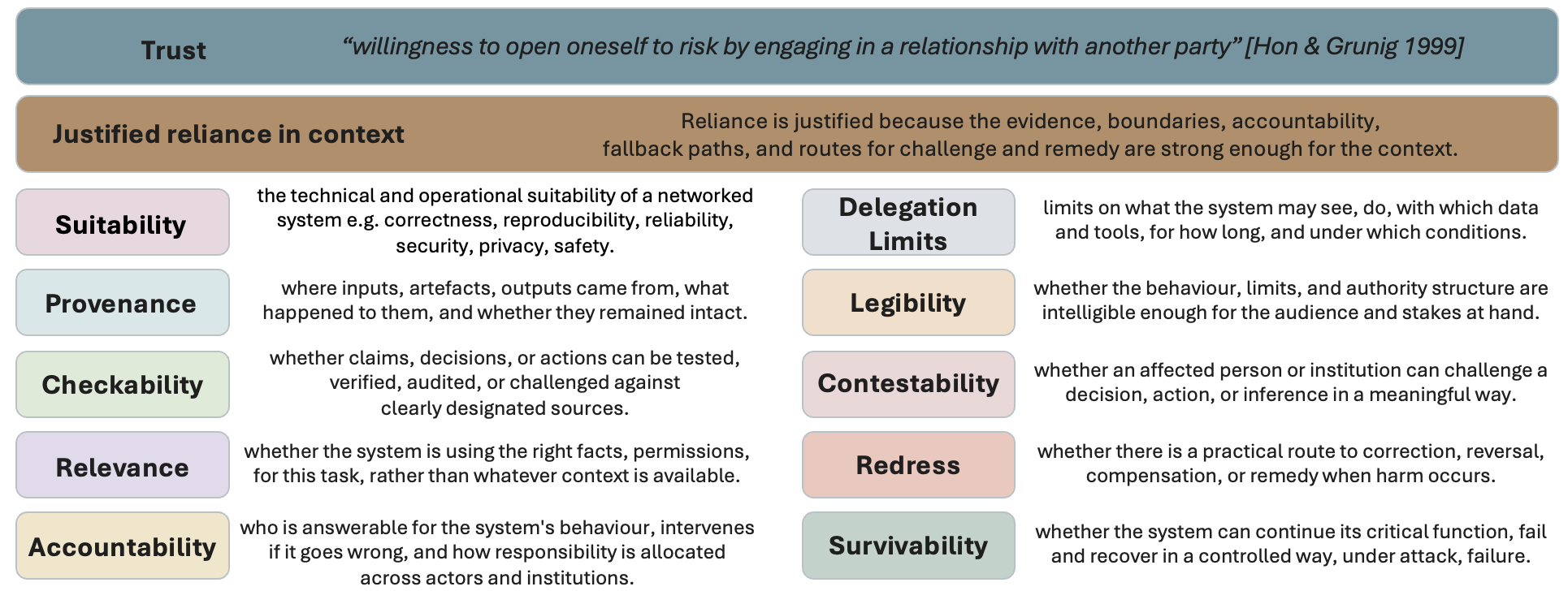}
\caption{Trust pillars \label{fig:trust-pillars}}
\end{figure}
\else
\begin{figure}[!h]
\centering
\includegraphics[width=.8\linewidth,height=.8\textheight,keepaspectratio]{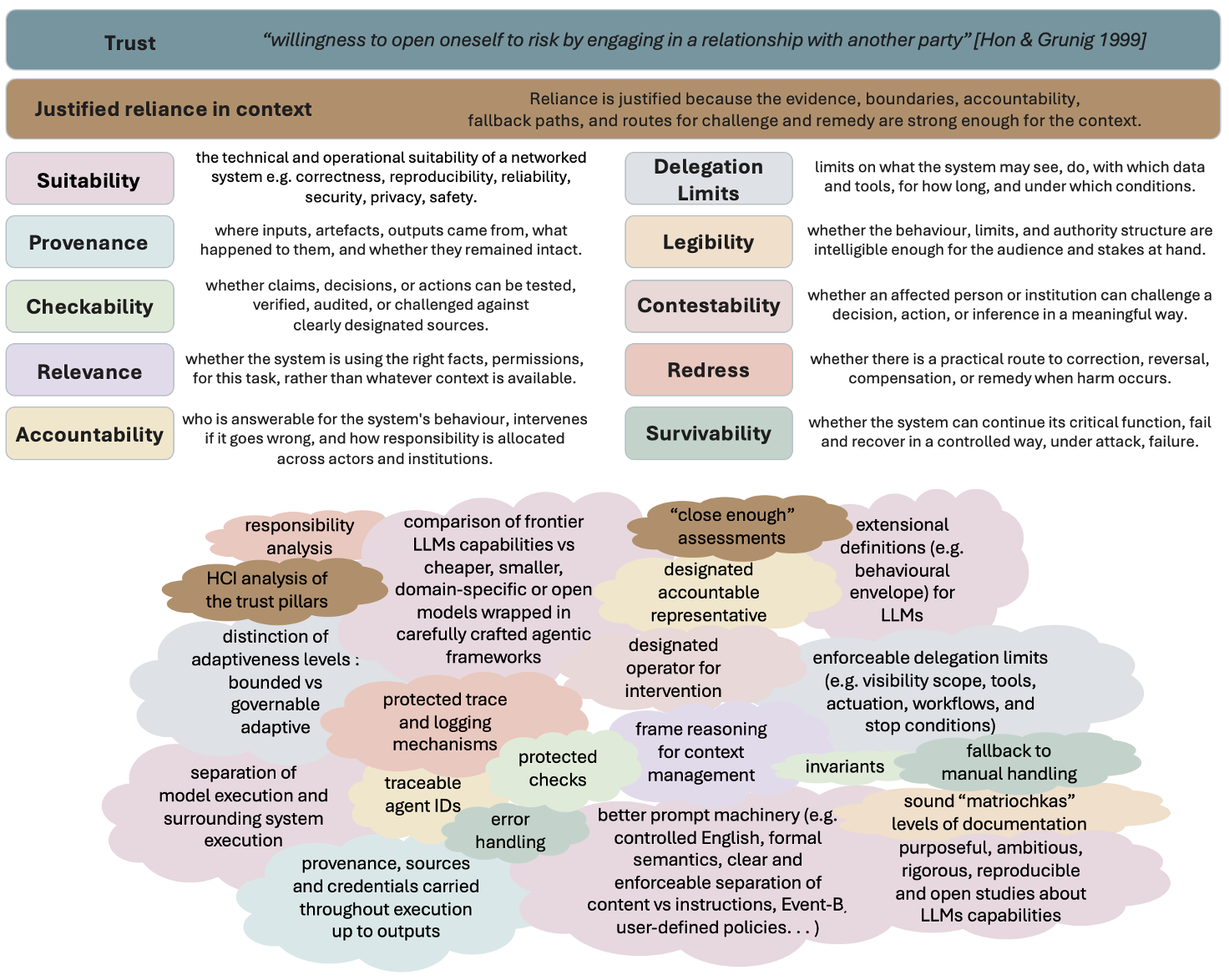}
\caption{Trust pillars and Suggestions \label{fig:trust-pillars}\label{fig:clouds}}
\end{figure}
\fi

For my grandma's passport, Suitability means e.g.  that her application reaches
the passport office when uploaded, and nobody else; Provenance: she can
demonstrate that she is the one who applied; Checkability: her application can
be checked w.r.t.  official requirements; Relevance: only the necessary
identity facts and images are in frame, nothing more; Accountability: the
passport office owns the process and intervention path; Delegation limits: no
silent widening into unrelated data collection or submission authority;
Legibility: the renewal state, e.g.  requirements and next steps, are
intelligible to users; Contestability: there exists a practically feasible
route to challenge rejection or misprocessing; Redress: there exists a
practically feasible path to seek correction to processing error, or submission
failure; Survivability: under uncertainty or system failure, we fall back
safely to manual handling.

There may be more pillars: I am not trying to argue that these are the only
ones, but I have tried for them not to be redundant. There may be fewer
pillars; arguably, we could join some pillars into more abstract ones~\ifarxiv---see
e.g.~\cite{ob17}\fi.  But I felt we needed relatively concrete pillars, hoping for
them to be actionable, perhaps even measurable, e.g. in engineering and civil
society: not all pillars can be addressed at a technical level, or not solely.
Redress for example needs legal and regulatory support, Accountability needs
organisational support.  Even within computing, this cannot be the prerogative
of certain groups, e.g. formal, safety or security folks, or be addressed by
one single company's solutions.

Tradeoffs between pillars may be envisageable in certain scenarios, and each
pillar needs discussing. For example, the Checkability pillar
mentions ``clearly designated sources'': if the system is deployed by, e.g.
the NHS or the CDC, they should clearly indicate what the reference sources
are. For a personal agent, users need the freedom to designate sources.
\ifarxiv
\begin{wrapfigure}{r}{0.31\linewidth} \centering \vspace{-1\baselineskip}
\includegraphics[width=1\linewidth]{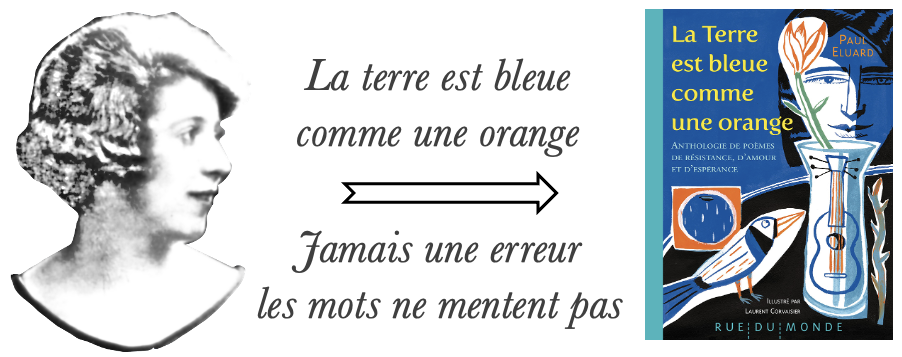} \label{fig:user-designated-sources} \vspace*{-11mm}
\end{wrapfigure}
\fi
I think it would be unwise, indeed likely at odds with trustworthiness, to impose
``sources of truth''. 
Rather the system needs to demonstrably preserve
provenance for reference material, and accept and demonstrate compliance with
user feedback w.r.t. more or less preferred sources. A source may or may not
coincide with fact; it may be poetry.  

\section{What agentic systems are we looking to trust?}

How much freedom to expand their remit may these systems have?  I found it
useful to define two levels: bounded agentic systems, and governable adaptive
agentic systems.  In practice, both levels may be distributed across device,
edge, cloud, and boundedness and governability must hold across execution
landscapes.

\emph{Bounded agentic systems} may plan, delegate, and act, but only inside
fixed bounds, e.g. visibility scope, tools, actuation, and stop conditions are
bounded in advance.  Delegation limits prevent silently widening permissions
transitively, across handoffs. Role adaptation is not allowed: executors remain
bounded executors; they do not alter the visibility scope, call new tools,
become new orchestrators, or open new workflows outside the fixed bounds.

This is hard today as LLMs can disregard their instructions, and the
distinction between actual content and instructions does not seem apparent.  So
we may need, amongst other things, better prompt machinery (e.g. controlled
English, formal semantics, clear and enforceable separation of content vs
instructions, Event-B, user-defined policies\ldots).

\emph{Governable adaptive agentic systems} generalise the bounded case: they
may adapt beyond fixed bounds, to delegation limits given by explicit
governance rules (e.g. a declarative description of the envelope of legitimate
behaviours, or invariants): visibility scope, tools, actuation, and stop
conditions are not bounded in advance. Role adaptation is allowed: an executor
may become a sub-orchestrator in a wider execution graph, but that adaptation
must stay within the envelope of behaviours given by the governance rules, with
explicit delegation rather than silent widening.

\emph{The trust pillars must be preserved throughout adaptation} and must
remain effective and inspectable: suitability, provenance, checkability,
relevance, accountability, delegation limits, legibility, contestability,
redress and survivability.

\section{How can we architect such systems?}
\ifarxiv
\begin{wrapfigure}{r}{0.31\linewidth} \centering \vspace*{-5mm}
\includegraphics[width=1\linewidth]{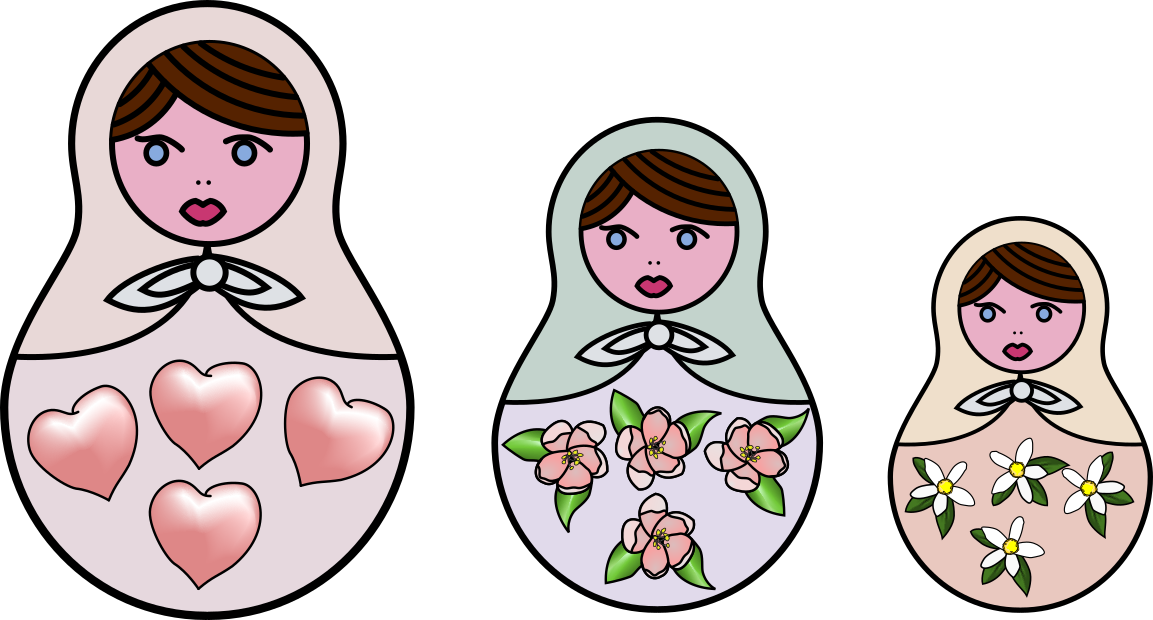} \label{fig:matriochkas} \vspace*{-9mm}
\end{wrapfigure}
\fi
Suitability and Delegation limits may be helped by existing mechanisms e.g.
identity and access management, sandboxing, attestations, protected storage,
logging, network controls, certification, computer architecture
foundations\ifarxiv~\cite{arm-cca}\fi\ldots C2PA’s content credentials~\cite{c2pa} may
help with Provenance. 
Legibility may be helped by using LLMs to turn complete, formal specifications
into more palatable yet sound documents, targetted at different
audiences---e.g. implementers, users, regulators, policy-makers---each level of
documentation soundly abstracting the original one, nested like matriochkas.
I am certain there's much more: a few suggestions appear in
Figure~\ref{fig:clouds}.  

\ifarxiv
\begin{figure}[!h] 
\ifcacm
\vspace*{-4mm}
\fi
\centering
\includegraphics[width=0.8\linewidth]{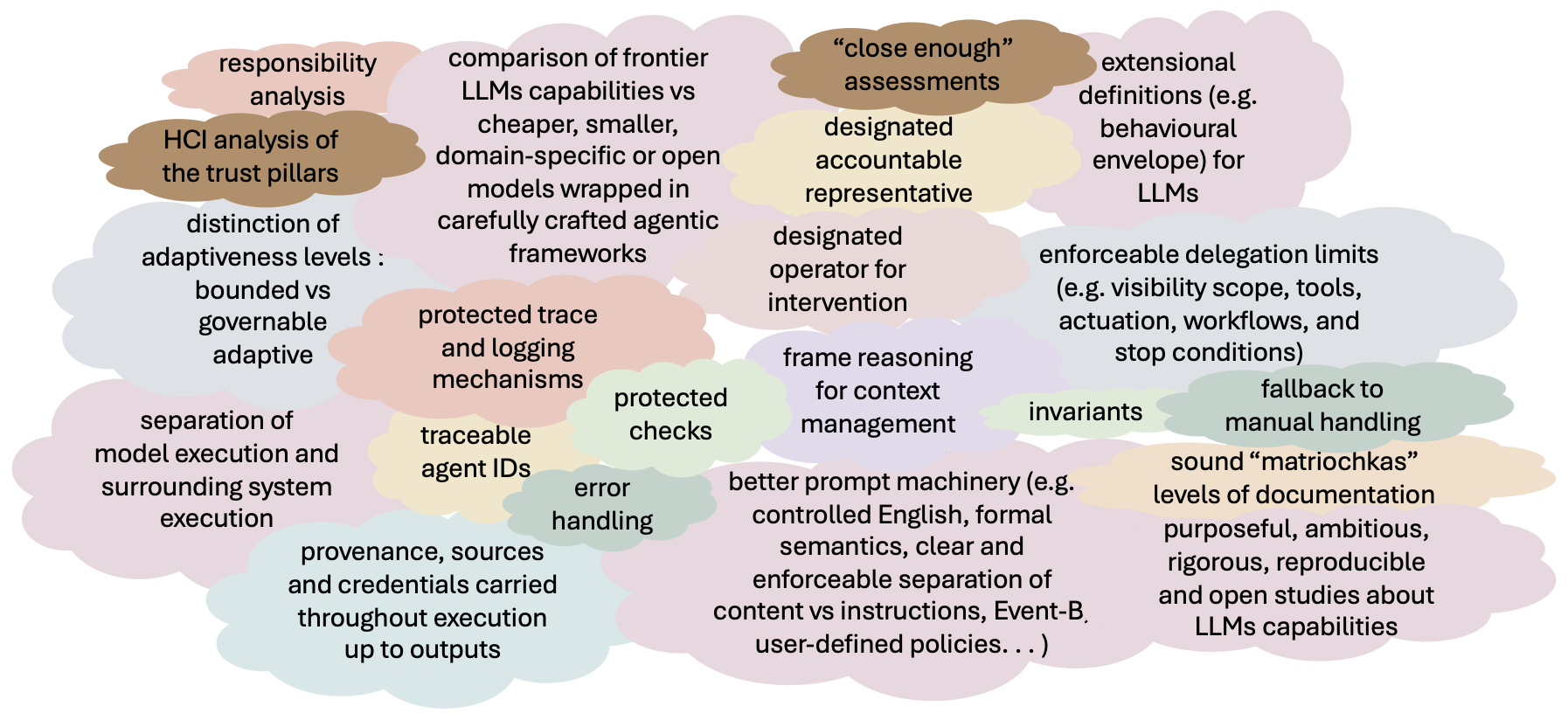}
\ifcacm
\vspace*{-7mm}
\else
\vspace*{-4mm}
\fi
\caption{A few suggestions \label{fig:clouds}}
\ifcacm
\vspace*{-5mm}
\fi
\end{figure}
\fi

I found inspiration for these whilst reflecting on
what trust might mean for users like my grandma in the context of agentic AI,
imagining what mechanisms might help meet the trust pillars. To let us imagine
further, here are a few papers that have resonated with me:
\begin{itemize}
\ifarxiv\item Adabara et al~\cite{survey-adabara} review agentic AI across
``architectural paradigms, threat taxonomies, and governance strategies''. They
emphasise the need for ``interdisciplinary collaboration'' and ``cooperative
innovation''.\fi
\item Chavali, Johnson and Denning~\cite{cjd26} examine regulations across
``the European Union, the United States, China, India, and Canada'', and
propose a ``Geneva Convention for AI'' to ``co-create
AI systems that adhere to ethical principles like transparency, accountability,
and professional oversight to promote societal well-being''.
\item Engin and Hand~\cite{engin-hand} propose a governance framework for AI
oversight, following three  dimensions: authority, autonomy, and
accountability. The framework aims to support ``the development of adaptive
regulations that can evolve alongside advancing AI capabilities without
compromising safety or accountability.''
\item Osmond and Jego~\cite{oj26} survey ``how public bodies,
international organisations, national regulators, and the private sector define
agentic artificial intelligence'', suggesting ``persistent definitional gaps''.
\item Shapira et al~\cite{baulab} deployed agents over $2$ weeks,
and observed ``unauthorized compliance with non-owners'', ``disclosure of
sensitive information'', ``uncontrolled resource consumption'', ``identity
spoofing'', ``cross-agent propagation of unsafe practices'', raising questions
of e.g. accountability and delegated authority.
\ifarxiv\item Shavit et al~\cite{openai-agentic} examine which parties are involved in
an agentic AI system lifecycle, and suggest best practices for each of these
parties for keeping agents’ operations safe and accountable. They list open
questions on how to implement these practices, and categories of impact from
agentic systems.\fi
\item Shen et al~\cite{shen} assess two public sector procurement checklists to
identify innovation directions for creating ``regulatable'' AI systems.  They
indicate that ``new methods are needed to connect data quality and objective
functions to outcomes, to identify what is possible with limited data and model
access, to monitor continuously learning agents, to balance transparency and
privacy, and to enhance effective human+AI interaction.''
\ifarxiv\item Wischnewski, Kr\"amer and M\"uller~\cite{wkm23} state that ``to achieve appropriate reliance, users’ trust should be calibrated to reflect a system’s capabilities.'', and review the literature across ``calibration'' dimensions: ``initial versus dynamic'', ``warranted versus unwarranted'', ``static versus adaptive'', and ``capabilities versus process-oriented''.\fi
\end{itemize}

\section{Would this be enough for my grandma?}

Trust is granted by the user. Therefore all we can do is try to give the user
justifications for relying on the system. These justifications will depend on
the user's context, background, preferences, and beliefs, but I have tried to list
a number of justification elements, or trust pillars, which might be just about
concrete enough to be actionable or measurable.
 
We need to understand our capability of control, or lack thereof, at various
adaptiveness levels: controlling bounded systems already seems hard with
state-of-the-art technology, and governable adaptive systems seem harder still.
Agentic systems likely need to evolve from prompt-only control, towards
additional enforceable architectural controls.  

What may the system see or do? How much can it delegate, or adapt?  What is
checked before action? Who is answerable?  Can the user challenge, correct, or
stop the system? If such questions remain implicit, my grandma should not
trust the system. If they are explicit, addressed by deliberate architecture,
perhaps trust becomes possible.

\ifarxiv
\par\medskip\noindent\textit{Acknowledgements.}\enspace
\else
\paragraph{Acknowledgements}
\fi
The opinions expressed in this column are my own and not necessarily those of
my employers.

Conversations with Eric Biscondi, Sylvan Clebsch, Hugo Ferreira, Hrutvik
Kanabar, Yuxi Liu, Nikos Nikoleris, Peter O'Hearn, Matthew Parkinson, Thomas
Speier and Jules Villard helped me gather my thoughts. Patrick Cousot inspired
me to look at Event-B and reviewed various drafts. Richard Grisenthwaite too
gave me helpful reviews, and suggested the image of Fort Knox vs hotel safe.
Rachel Coldicutt pointed me to the O'Neill reference, and discussed cultural
aspects of AI-assisted farming.  Daryl Stewart drew the ``Oma'' and ``Opa''
love hearts, the ``Abstraction Matriochkas'', and the ``Granny thinks of poetry''
picture, whilst giving me meaningful feedback.  

Finally, the East London Massive \& Co. lost a dear friend and mentor earlier
this year,  Richard Bornat. I would have loved (truly) to hear his thoughts on
this column; but this was not to be.

\bibliographystyle{plain}
\bibliography{trust}

\end{document}